# Room-temperature polariton lasing in quantum heterostructure nanocavities

Jang-Won Kang[1], Bokyung Song[1], Wenjing Liu[3], Seong-Ju Park[2], Ritesh Agarwal[3]*, Chang-Hee Cho[1]*

[1]Department of Emerging Materials Science, Daegu Gyeongbuk Institute of Science and Technology (DGIST), Daegu 42988, South Korea

[2]School of Materials Science and Engineering, Gwangju Institute of Science and Technology, Gwangju 61005, South Korea

[3]Department of Materials Science and Engineering, University of Pennsylvania, Philadelphia, Pennsylvania 19104, USA

*e-mail: chcho@dgist.ac.kr (C.H.C.); riteshag@seas.upenn.edu (R.A.)

**Controlling light-matter interactions in solid-state systems has motivated intense research to produce bosonic quasi-particles known as exciton-polaritons, which requires strong coupling between excitons and cavity photons. Ultra-low threshold coherent light emitters can be achieved through lasing from exciton-polariton condensates, but this generally requires sophisticated device structures and cryogenic temperatures. Polaritonic nanolasers operating at room temperature lie on the crucial path of related research, not only for the exploration of polariton physics at the nanoscale but also for potential applications in quantum information systems, all-optical logic gates, and ultra-low threshold lasers. However, at present, progress toward room-temperature polariton nanolasers has been limited by the thermal instability of excitons and the inherently low quality factors of nanocavities. Here, we demonstrate room-temperature polaritonic nanolasers by designing wide-gap semiconductor heterostructure nanocavities to produce thermally stable excitons coupled with nanocavity photons. The resulting mixed states of exciton-polaritons with Rabi frequencies of approximately 370 meV enable persistent polariton lasing up to room temperature, facilitating the realization of miniaturized and integrated polariton systems.**

Lasing from exciton-polariton condensates in solid-state systems is distinguished from conventional lasing by the exhibition of low-threshold polariton lasing with macroscopic and spontaneous coherence (1–5). Strong coupling between excitons and photons is crucial for the formation of the bosonic quasi-particles known as exciton-polaritons, which is generally achieved by means of two-dimensional semiconductor quantum wells (QWs) embedded in high-finesse optical microcavities (1,4,5). Despite the design of elaborate structures to obtain high-quality excitons and cavity photons, polariton lasing has mostly been observed at cryogenic temperatures due to the small binding energies (~10 meV) of Wannier-Mott excitons in typical semiconductors (4,5). With wide-gap semiconductors of larger exciton binding energies such as GaN (~20 meV) and ZnO (~60 meV), the room-temperature polariton lasing has been recently achieved under both the optical (6–8) and the electrical injections (9). Moreover, organic (10) and polymer (11) materials have also showed the polariton condensation and lasing at room temperature owing to the large binding energy (~1 eV) of Frenkel excitons. However, these devices have required the complicated microcavity structures (6–11).

The coupling strength between excitons and cavity photons scales as $\Omega \sim \sqrt{(f/V_m)}$, where $\Omega$ is the Rabi frequency, $f$ is the oscillator strength of the excitons, and $V_m$ is the mode volume of the cavities (3). Semiconducting nanostructures such as nanorods simultaneously produce excitons and cavity photons with reduced mode volumes, boosting their interaction time, and thus, the coupling strength. Indeed, size-tunable and enhanced coupling strengths have been observed in semiconductor nanowire cavities, enabling strong coupling without high-finesse microcavities (12,13). The development of room-temperature polaritonic nanolasers, however, is still limited by the thermal instability of the excitons and the inherently low quality factors of the nanocavities (12,14).

Here, we demonstrate room-temperature polaritonic nanolasers of a simple nanorod geometry by designing wide-gap semiconductor heterostructure nanocavities, each composed of a nanorod core enclosed by a radial QW shell. The radial QW structure produces thermally stable excitons with an enhanced oscillator strength, which enables strong coupling with photons confined in small mode volume (0.1 μm$^3$) nanocavity, giving rise to Rabi frequencies of ~370 meV at room temperature. Our unique approach exploits these quantum heterostructure nanocavities to enable persistent polariton lasing up to room temperature. Furthermore, a double threshold behavior well below and above Mott density, observed in the quantum heterostructure nanocavities, evidences the transition from polariton to photon lasing regimes.

The quantum heterostructure nanocavity used in this study consists of a ZnO nanorod core and a multi-QW shell with five pairs of 4-nm-thick ZnO QW layer and 10-nm-thick $Zn_{0.9}Mg_{0.1}O$ barrier layer, as shown schematically in Fig. 1a. The radial QW structure is grown on an array of ZnO cores by means of metal-organic chemical vapor deposition (MOCVD) (see Materials and Methods). Figure 1b presents a scanning electron microscope (SEM) image of an as-grown QW nanorod array, with an average rod diameter of ~600 nm and a rod length of ~2.7 μm, which belongs to a waveguide structure to support Fabry-Pérot cavity modes along the axial direction over the emission wavelength range of ZnO QWs. The conformal growth of single-crystalline radial QWs is confirmed by radial cross-sectional scanning transmission electron microscope (STEM) images (Fig. 1c and 1d), which show the 5-period QW and barrier layers. Note that the QW nanorod depicted in Fig. 1c and 1d was encapsulated with an additional ZnO layer to obtain high-contrast STEM images of the QW region. Structural analysis via axial cross-sectional energy-dispersive X-ray spectroscopy, as shown in Fig. 1e, further confirms the highly conformal growth of the QW layers. To investigate the exciton properties of the QW nanorods, spatially resolved micro-photoluminescence (PL) measurements were performed on single QW nanorods transferred onto a SiO$_2$/Si substrate. Figure 1f presents the exciton luminescence spectrum of a single ZnO QW nanorod in comparison with that of a bare ZnO nanorod (without QWs), where each spectrum was measured at the center of the nanorod under continuous-wave 325 nm laser excitation with a low pump power density of 0.67 kW/cm$^2$ at room temperature. Compared with the bare nanorod, which exhibits an emission peak at ~3.26 eV with a full-width at half-maximum (FWHM) of 88 meV, the QW nanorod shows an exciton emission intensity that is much stronger, by a factor of more than 10, and peaks at ~3.32 eV with a FWHM of 58 meV. The observed PL peak (~3.26 eV) for the bare nanorod corresponds to the first longitudinal-optical (LO) phonon replica of free A exciton (3.306 eV) due to strong exciton-LO phonon coupling in the bulk ZnO (*15*). However, the PL peak (~3.32 eV) for the quantum well nanorod is primarily dominated by zero-phonon excitonic emission because the coupling between exciton and LO phonon is greatly reduced in QW structures (*16,17*). The observed blue-shift, enhanced intensity, and reduced broadening of the exciton emission peak from the QW nanorod originate from quantum confinement effects (*18*), indicating that the radial QW structure provides thermally stable excitons with an enhanced oscillator strength even at room temperature (Supplementary Materials S1).

The formation of exciton-polaritons and their coupling strengths in waveguide geometries can be analyzed by measuring the waveguided PL at one end of a nanorod when the other end of the nanorod is excited with a focused laser spot (see Methods). Note that a low level of excitation (0.67 kW/cm$^2$) was used to create an exciton density ($8.8 \times 10^{16}$ cm$^{-3}$) well below the Mott density ($1.5 \times 10^{18}$ cm$^{-3}$) because above the Mott density, the excitons dissociate into an electron-hole plasma (*14,19*). Strong coupling between the excitons and cavity photons in a QW nanorod gives rise to coupled eigenstates consisting of upper and lower polariton branches with a vacuum Rabi frequency ($\Omega$) (*20*), which is the frequency of the oscillation between the exciton and photon states. Figure 2a presents the PL spectra at room temperature, measured at the end and the body (center) of a QW nanorod with a diameter of 640 nm and a length of 2.75 μm. The body emission is dominated by a Lorentzian-shaped single peak centered at ~3.32 eV, which is attributed to the emission from QW excitons. In contrast, the end emission is composed of the weak shoulder peak at ~3.32 eV on the higher-energy side and the strong emission at ~3.20 eV on the lower-energy side. The weak shoulder peak of the end emission, which corresponds to the body emission peak, results from the detection of uncoupled-exciton emission due to the short distance between excitation and detection spots. The strong emission centered at ~3.20 eV on the lower-energy side is assigned to the emission from the lower polariton branch. Importantly, the strong and broad emission from the lower polariton branch reveals clear multiple resonance peaks, resulting from the eigenmodes of the exciton-polaritons, which are composite quasi-particles consisting of the QW excitons and the Fabry-Pérot cavity photons in the QW nanorod. By fitting the measured resonance peaks to the theoretical model for one-dimensional exciton-polariton waveguide modes, the energy−momentum ($E$–$k_z$) dispersion of the exciton-polaritons along the long axis of the ZnO QW nanorod can be obtained (*13,21*). Figure 2b presents the $E$–$k_z$ dispersion of the exciton-polaritons for a ZnO QW nanorod with the diameter of 640 nm and length of 2.75 μm. Note that the data points represent measured resonance energies with equidistant momenta at integer values of $\pi/L_z$, where $L_z$ is the length of the nanorod. To determine the dominant cavity modes that are valid for the observed exciton-polariton eigenmodes, the electromagnetic fields in the nanorod cavity were numerically calculated using a finite-difference time-domain (FDTD) method (Supplementary Materials S2). The calculated results indicate that the dominant cavity modes in the QW nanorod are $HE_{22}$ and $HE_{13}$. Then, the $E$–$k_z$ relation was calculated using the theoretical model for one-dimensional (1D) exciton-polariton waveguide dispersion (Supplementary Materials S3). The experimentally measured resonance energies with equidistant momenta at the interval of $\pi/L_z$ were fitted to the calculated $E$–$k_z$ polariton dispersion by a parallel shift of the data points in the momentum axis. Note that the initial values of the equidistant momenta can be roughly estimated by simulating the field distribution of the Fabry-Pérot cavity mode since $k_z = n\pi/L_z$ where *n* is the number of antinodes along the *z*-axis (Supplementary Materials S2). The oscillator strength was also adjusted as a fit parameter. Finally, an excellent fit was obtained with the enhanced oscillator strength of the QW excitons by a factor of 11 compared to the bulk excitons when the QW excitons are coupled to the HE$_{22}$ cavity mode (inset of Fig. 2b). The $E$–$k_z$ dispersion of the exciton-polaritons for the ZnO QW nanorod confirms the strong coupling with a Rabi splitting energy ($\hbar\Omega$) of ~370 meV.

Such large Rabi oscillator strength ($\hbar\Omega$) in the QW nanorods results from the enhanced oscillator strength of QW exciton and the reduced mode volume of nanocavity (*3,22*). Temperature-dependent PL measurements reveal that the emission intensity of the QW nanorod is about 10 to 20 times higher than that of the bare nanorod over the temperature range of 77–297 K, while the activation energy for non-radiative process is almost the same for both the QW (26 meV) and bare nanorods (23 meV) (Supplementary Information S1) (*23*). This indicates that the oscillator strength is greatly enhanced for the QW exciton by a factor of more than 10, while the non-radiative rates remain almost same (Supplementary Materials S1) (*24*). Taking into account the enhanced oscillator strength (by a factor of 11), the mode volume (0.1 μm$^{-3}$), and the spatial overlap (0.23) between the field and the QWs, the Rabi splitting energy ($\hbar\Omega$) was calculated to be about 360 meV, which is in excellent agreement with the value obtained from the polariton dispersion (Supplementary Materials S4).

To investigate the lasing characteristics of ZnO QW nanorods at room temperature, micro-PL measurements were conducted for individual QW nanorods under 355 nm pulsed excitation at various pump fluences (see Materials and Methods). Figure 3a shows an SEM image of a single ZnO QW nanorod transferred onto a SiO$_2$/Si substrate. Figures 3b–e present optical images showing a transition from spontaneous to coherent emission with increasing pump fluence. Above the lasing threshold, the typical interference patterns associated with longitudinal Fabry-Pérot cavity modes are observed from the QW nanorod (*25*). The interference pattern results from the coherent light emission from the two end facets of the nanorod, at which the light is emitted nearly spherically due to diffraction at the nanoscale apertures (*25*). This coherence in light emission is further evidenced by angle-resolved emission spectra showing the phase-correlated interference patterns (*26*), resembling Young's double-slit experiments, where the coherent light emitted from the two end facets of the nanorod interferes with each other (Supplementary Materials S5). Figure 3f presents the micro-PL spectra of the ZnO QW nanorod at room temperature. At a pump fluence of 286 μJ/cm$^2$, an abrupt increase in emission intensity is observed at the lasing peak energy of 3.225 eV, whereas broad spontaneous emission is observed below the lasing threshold. A remarkable feature is that the single-mode lasing near the threshold has an extremely narrow linewidth of 0.7 meV even at room temperature. Figure 3g shows the spectral map of the lasing behavior as a function of the pump fluence. A gradual blue-shift of the single-mode lasing peak is observed with increasing pump fluence at the range below the Mott density (< 810 μJ/cm$^2$), which results from the repulsive interactions between the polariton and the uncondensed exciton reservoir, and also between the polaritons (*27*). As the pump fluence is further increased above the Mott density (> 810 μJ/cm$^2$), the system enters a weak coupling regime, another lasing peak centered at ~3.195 eV emerges on the lower-energy side, which can be explained by the shift of gain regime and relevant mode change to the lower energy side due to the band-gap renormalization at this higher level of excitation (*28*). With the gradual transition from excitons to electron-hole plasma, the additional blue-shift of lasing peak occurs due to the reduction of exciton binding energy and change of refractive index (*28*). Figure 3h compares the lasing spectra at low (360 μJ/cm$^2$) and high (2272 μJ/cm$^2$) pump fluences, showing a marked change in the lasing mechanism as the pump fluence increases.

Polariton lasing essentially requires excitons to be strongly coupled with photons under above-threshold excitation. However, as the excitation increases, the exciton density also increases, and Coulomb screening reduces the binding energy of the excitons, causing the system to eventually enter the regime of an uncorrelated electron-hole plasma. The crucial density criterion separating the excitonic and electron-hole plasma regimes is the Mott density, where the experimentally measured value in bulk ZnO is $1.5 \times 10^{18}$ cm$^{-3}$ at room temperature (*19,29*). For an ideal 2D QW, the exciton Bohr radius is decreased to a half that of the equivalent 3D exciton (*30*), leading to higher Mott density for the 2D excitons. However, for the ZnO QWs used in this study, the Mott density would be slightly larger or similar with the value for bulk ZnO because the thickness of QW (4 nm) is larger than the exciton Bohr radius of bulk ZnO (~1.8 nm) (*14*). Indeed, the Mott densities in realistic GaN and GaAs QWs are almost the same with those of the bulk counterparts (*31–34*). Thus, the Mott density of bulk ZnO is used for both the QW and bare nanorods to qualitatively compare their lasing characteristics.

To verify that the observed lasing action is due to the coherence of the exciton-polaritons in the ZnO QW nanorods, we investigated the lasing characteristics with increasing electron-hole pair (EHP) densities by varying the excitation fluence. Figure 4a shows a double-logarithmic plot of the integrated intensity as a function of the incident pump fluence at room temperature, and the results confirm the clear threshold behavior of the lasing phenomenon. Notably, the ZnO QW nanorods exhibit a lasing threshold of 286 μJ/cm$^2$, which is much lower than the critical transition excitation of 810 μJ/cm$^2$ corresponding to the Mott density. Above the lasing threshold, the polariton dispersion is still maintained with a Rabi splitting energy ($\hbar\Omega$) of ~360 meV, indicating the polariton lasing in strong coupling regime (Supplementary Materials S6). Moreover, the QW nanorods show behavior characteristic of a second threshold above the critical excitation corresponding to the Mott density, where a transition from polariton to photon lasing occurs. This second threshold behavior evidences the transition from strong to weak coupling regime with the onset of photon lasing, as observed in GaAs-based low-dimensional microcavities (*35,36*). Together with the observed blue-shift of the lasing peak (Fig. 3g), the second threshold behavior strongly supports that the lasing observed in the ZnO QW nanorods originates from strong coupling of the exciton-polaritons.

The double threshold behavior for the QW nanorods was observed reproducibly, as provided in the statistical data (Supplementary Materials S7).

The same plot for bare ZnO nanorods is also displayed for comparison. A stark contrast can be seen from the fact that the lasing threshold of the bare ZnO nanorods appears at an excitation (905 µJ/cm$^2$) that is above the critical value corresponding to the Mott density, meaning that the lasing in the bare nanorods is photon lasing from the electron-hole plasma (Supplementary Materials S8). Statistics on the lasing threshold for the bare nanorods are also provided in Supplementary Materials S7. Figure 4b presents a double-logarithmic plot of the linewidth versus the pump fluence. For both the QW and bare nanorods, the lasing linewidths are well below the linewidth associated with the calculated quality factor (< 500) for the photonic cavity modes in the nanorod geometry. Important distinction is that for the QW nanorods, the sharp linewidth is maintained until the Mott density is reached, and the linewidth then gradually increases as the system enters the photon lasing regime, whereas for the bare nanorods, a simple increase is seen in the linewidth of the lasing peak due to the change in the refractive index associated with the electron-hole plasma (*37*). To further investigate the lasing behavior in relation to the excitation, we measured the EHP density at the threshold, as obtained from the measured threshold pump fluence, in the temperature range from 77 to 297 K. Note that the threshold EHP density per QW was estimated using the net absorption in the QW structure calculated via transfer matrix methods (Supplementary Materials S9). Figure 4c plots the threshold EHP density as a function of temperature for both the QW nanorods and the bare nanorods. For the QW nanorods, the threshold EHP density is as low as $4.5 \times 10^{16}$ cm$^{-3}$ at 77 K and slowly increases with increasing temperature, up to $2.8 \times 10^{17}$ cm$^{-3}$ at room temperature (297 K); all of the values over the entire temperature range are well below the Mott density of $1.5 \times 10^{18}$ cm$^{-3}$. By contrast, the bare nanorods show a much more rapid increase in the threshold EHP density with increasing temperature; the threshold density eventually exceeds the Mott density above 257 K and reaches a value of $3.2 \times 10^{18}$ cm$^{-3}$ at room temperature. Figure 4d presents the temperature-dependent evolution of the lasing mode for the QW nanorods. The data show step-like decreases in the energy of the lasing mode with increasing temperature due to the discrete nature of the polariton eigenmodes, and the decreases in the energy of the lasing mode are closely correlated with the temperature-dependent variation in the exciton resonance. More importantly, the energy separation between the lasing mode and the exciton resonance corresponds to the LO phonon energy of ~72 meV (*14*), implying that the population of the polariton eigenmodes is dominated by longitudinal optical phonon relaxation from the exciton reservoir (*38*). By contrast, the lasing mode of the bare nanorods appears near or above the exciton resonance energy in the temperature range of 267–297 K as a result of the photon lasing in the electron-hole plasma regime (Supplementary Materials S10). These results highlight the crucial role of quantum heterostructures in the realization of room-temperature polariton lasing in single nanostructures.

We have successfully demonstrated room-temperature exciton-polariton lasing in quantum heterostructure nanocavities. By virtue of the thermal stability, enhanced oscillator strength, and strong coupling with cavity photons of the QW excitons, such a quantum heterostructure nanocavity shows persistent exciton-polariton lasing up to room temperature. The polariton lasing is confirmed with large Rabi splitting energies in one-dimensional exciton-polariton dispersion relations. The double threshold behavior well below and above Mott density, observed in the quantum heterostructure nanocavities, evidences the transition from polariton to photon lasing regimes. Our demonstration opens up the possibility of studying various intriguing polaritonic phenomena in nanoscale systems and enables the realization of efficient nanolasers, all-optical logic gates, and quantum computing devices operating at room temperature.

**Methods**

**Device fabrication.** Quantum heterostructure nanocavities were fabricated via a two-step growth procedure. First, ZnO nanorod arrays were synthesized using a patterned nano-hole photoresist mask on ZnO (200 nm)/sapphire substrates by means of a hydrothermal method using an aqueous solution of equimolar zinc nitrate hydrate and hexamethylene-tetramine. Then, five pairs of ZnMgO/ZnO layers were grown on the core ZnO nanorods using a shower-head-type MOCVD reactor equipped with computer-controlled gas flow systems. Diethylzinc (DEZn), bis (cyclopentadienyl) magnesium (CP$_2$Mg), and O$_2$ gas were used as sources of Zn, Mg, and O, respectively. Ar was used as the carrier gas. Zn$_{0.9}$Mg$_{0.1}$O was grown at 810 °C and a CP$_2$Mg to DEZn flow ratio of 0.2. The growth conditions for high quality QW structures were optimized by controlling the (DEZn+CP$_2$Mg)/O$_2$ ratio.

**Optical measurements.** The as-grown nanocavities were dry-transferred onto 400 nm-thick SiO$_2$-coated Si substrates for optical measurements performed using a home-built microscope equipped with objectives (Nikon). The excitation power was adjusted using neutral density filters. For spatially resolved micro-PL spectroscopy, a continuous wave 325 nm He-Cd laser (Kimmon Koha) was focused by an objective (60×, 0.7 NA) to a beam spot size of ~3 μm with a power density range of 0.67 – 2.2 kW/cm$^2$, corresponding to an estimated excited EHP density of $8.8 \times 10^{16} - 3.5 \times 10^{17}$ cm$^{-3}$. PL spectra were collected using the same objective and an optical fiber on the focal image plane, resulting in a spatial resolution of less than 300 nm. The signal collected by the optical fiber was coupled to a 0.5 m spectrometer (Acton) and a cooled charge-coupled device (Pixis 2K, Princeton Instruments) with a spectral resolution of 0.1 nm. The lasing characteristics of the nanocavities were examined using the same optical setup but with excitation by a pulsed wave 355 nm laser (Teem Photonics) with a repetition rate of 1 kHz and a pulse width of 350 ps, which was focused by a 20×, 0.45 NA objective to a beam spot size of ~14 μm. Temperature-dependent measurements were conducted using a liquid-nitrogen-cooled cryostat (Janis Research).

**Acknowledgements**

This work was supported by the Basic Science Research Program (2016R1A2B4014448, 2016R1A6A3A11933287) and the Leading Foreign Research Institute Recruitment Program (2012K1A4A3053565) through the National Research Foundation of Korea, and by the DGIST R&D Program (17-BT-02) funded by the Ministry of Science and ICT of the Korean Government.

**Figures**

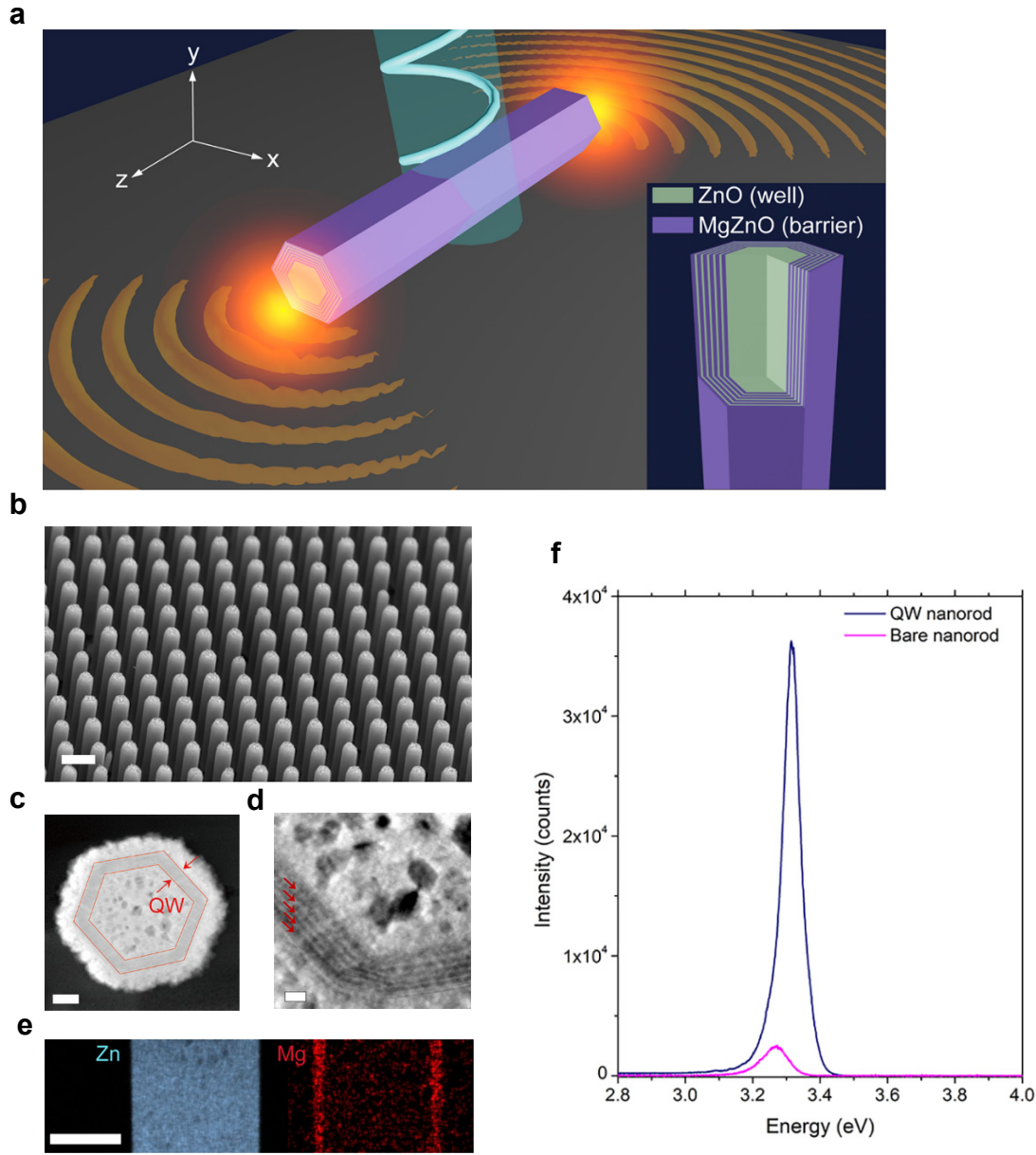

**Figure 1. Quantum heterostructure nanocavities.** (**a**) Schematic illustration of a single nanorod polariton laser on a SiO$_2$ substrate. The inset shows the shell structure, consisting of five pairs of Zn$_{0.9}$Mg$_{0.1}$O/ZnO QW layers. (**b**) SEM image of an as-grown QW nanorod array with an average rod diameter of ~600 nm and a rod length of ~2.7 μm. Scale bar: 1 μm. (**c**) STEM image showing a radial cross section of a QW nanorod. Scale bar: 100 nm. (**d**) Magnified STEM image of the QW region. The arrows indicate the ZnO wells. Scale bar: 20 nm. Note that the QW nanorod depicted in Fig. 1c and d was encapsulated with an additional ZnO layer to obtain high-contrast images of the QW region. (**e**) Elemental mapping images of an axial cross section of a QW nanorod, showing the conformal growth of QW layers on the core ZnO nanorod. Scale bar: 300 nm. (**f**) Exciton luminescence spectrum of a single QW nanorod at room temperature compared with the spectrum of a bare nanorod.

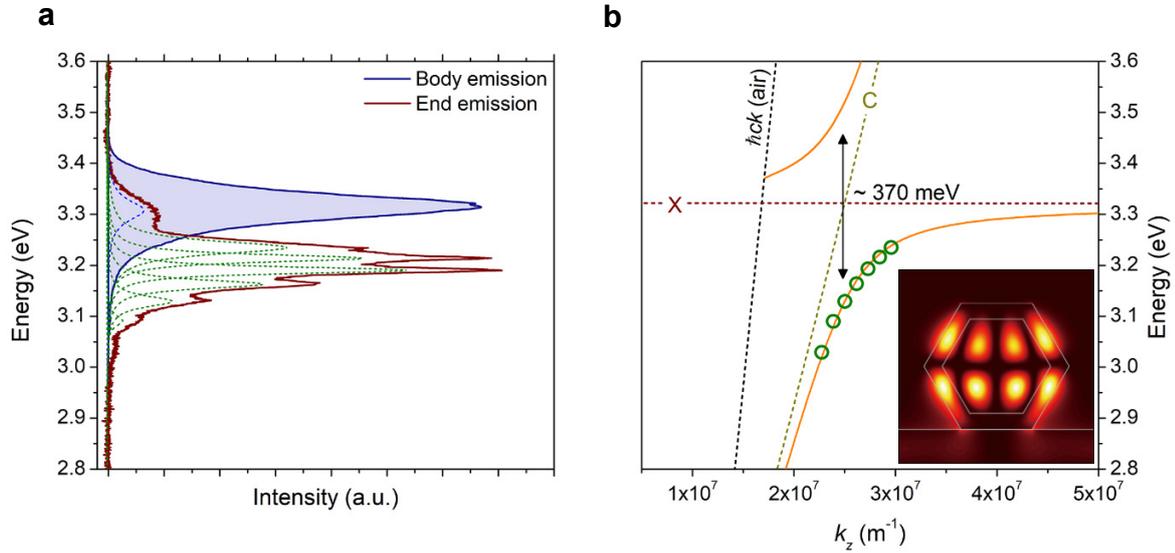

**Figure 2. Polariton dispersion at room temperature.** (**a**) Excitonic (blue line) and waveguided (red line) photoluminescence spectra from the body center and end of a QW nanorod (diameter: 640 nm, length: 2.75 μm), respectively. The end emission exhibits multiple resonance peaks, which correspond to the eigenmodes of the exciton-polaritons in the quantum heterostructure nanocavity. The dashed greeen lines represent the fitted Lorentzian line shapes used to determine the resonance energies in the lower polariton branch. (**b**) Dispersion relation of the exciton-polaritons along the long axis of the QW nanorod. The data points represent measured resonance energies with equidistant momenta at integer values of $\pi/L_z$, where $L_z$ is the length of the nanorod. The orange solid line represents the fit to the theoretical model for the exciton-polariton $HE_{22}$ guided mode, indicating an estimated Rabi splitting energy of ~370 meV. The horizontal red dashed line represents the QW exciton energy. The inset shows the calculated electric field intensity ($|E_y|^2$) profile in the radial cross-sectional plane for the $HE_{22}$ guided mode at an energy of 3.20 eV, where white lines indicate the QW region.

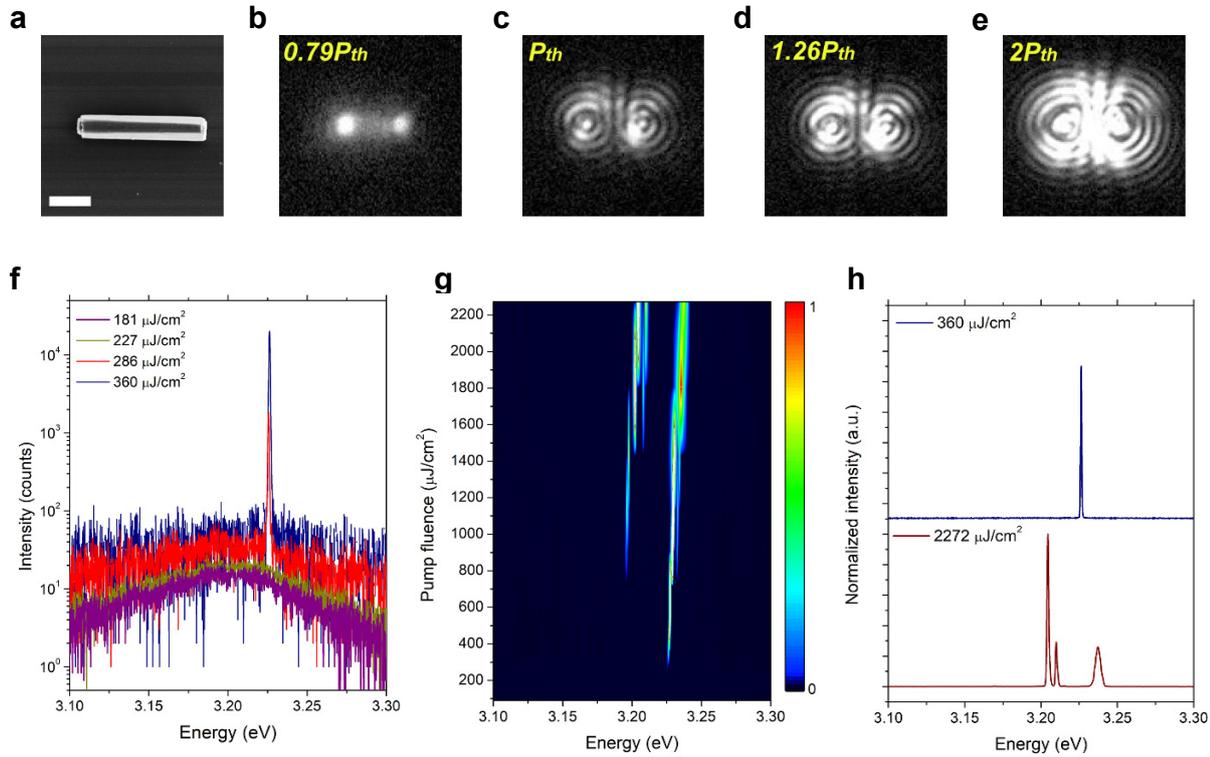

**Figure 3. Room-temperature polariton lasing spectra.** (**a**) SEM image of a single QW nanorod transferred onto a SiO$_2$/Si substrate. Scale bar: 1 μm. (**b–e**) Optical images showing a transition from spontaneous to coherent emission with increasing pump fluence. The interference patterns associated with longitudinal Fabry-Pérot cavity modes are observed. (**f**) Spectral evolution of a single QW nanorod laser (diameter of 580 nm, length of 2.70 μm) near the lasing threshold. (**g**) Spectral map of the lasing behavior of the QW nanorod laser as a function of pump fluence. (**h**) Normalized emission spectra of the QW nanorod laser at low (360 μJ/cm$^2$) and high (2272 μJ/cm$^2$) pump fluences.

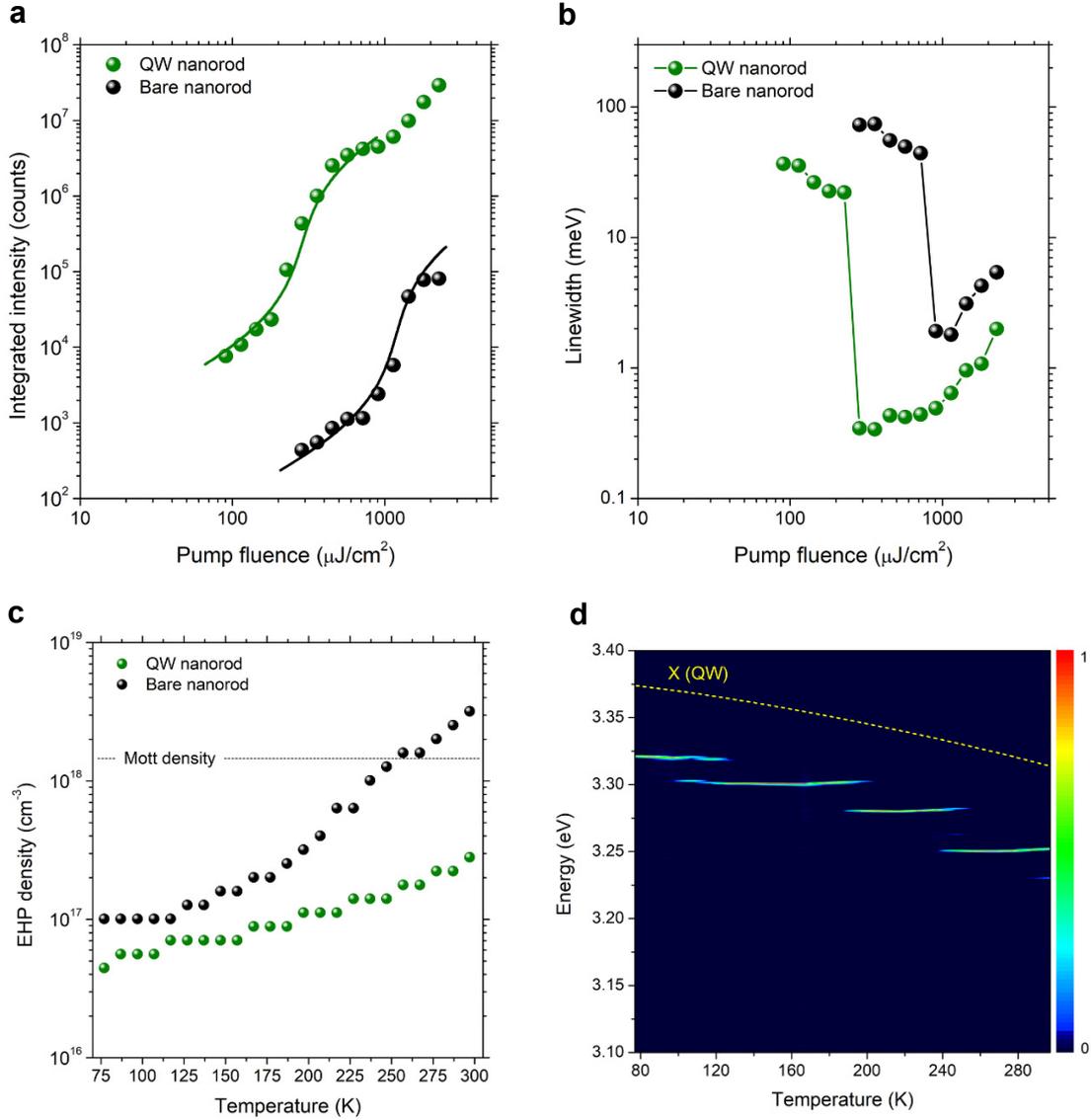

**Figure 4. Density and temperature dependence of polariton lasing.** (**a, b**) Double-logarithmic plots of integrated intensity (**a**) and linewidth (**b**) as a function of pump fluence at room temperature for a QW nanorod and a bare nanorod, showing the clear threshold behavior of the lasing. (**c**) EHP density at the threshold pump fluence in the temperature range from 77 to 297 K for both the QW and bare nanorods. The horizontal dotted line indicates the Mott density of ZnO at room temperature. (**d**) Spectral map of the QW nanorod laser measured at near-threshold fluences ($P_{th} \leq P \leq 1.26 P_{th}$) in the temperature range from 77 to 297 K, showing the temperature-dependent evolution of the lasing mode. The yellow dotted line represents the temperature-dependent exciton energy of the QW nanorod.

# Supporting Information

**Room-temperature polariton lasing in quantum heterostructure nanocavities**

Jang-Won Kang, Bokyung Song, Wenjing Liu, Seong-Ju Park, Ritesh Agarwal*, & Chang-Hee Cho*

**S1. Temperature-dependent exciton luminescence**

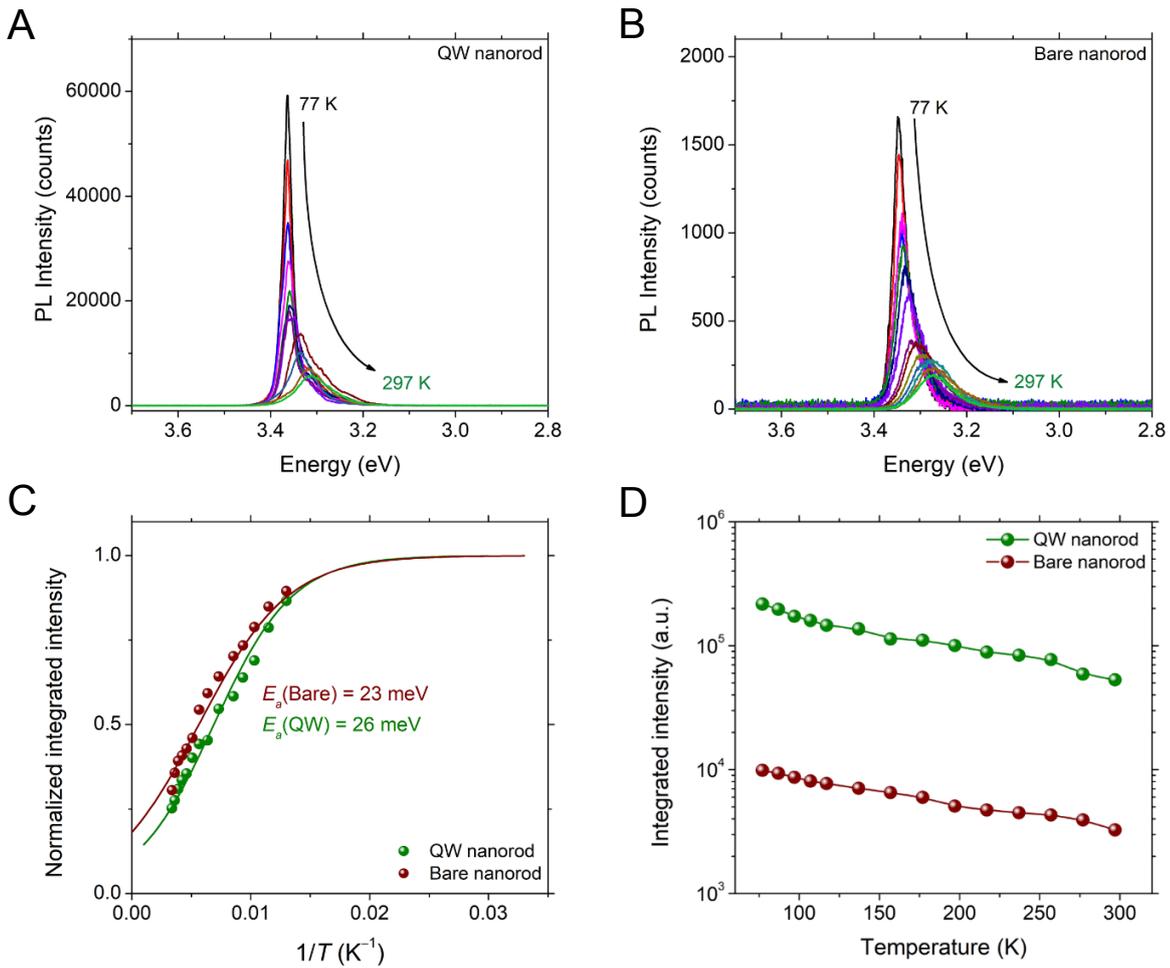

**Figure S1**. **Temperature-dependent photoluminescence characteristics in QW and bare nanorod cavities.** (**A, B**) Photoluminescence spectra of QW (A) and bare (B) nanorods at the temperature range from 77 to 297 K. (**C**) Arrhenius plot of integrated intensity for QW and bare nanorods. An activation

energy of non-radiative recombination was fitted by using the equation of $I(T) = [1+C exp(-E_a/kT)]^{-1}$ (*C*: constant, $E_a$: activation energy of non-radiative recombination) (*1*). The activation energy ($E_a$) of non-radiative recombination was estimated to be 26 and 23 meV for QW and bare nanorods, respectively, indicating that the activation energy for non-radiative process is almost the same for both the QW and bare nanorods. (**D**) The integrated intensity of QW and bare nanorods at the temperature range from 77 to 297 K, showing that the emission intensity of QW nanorod is increased by a factor of more than 10, compared to that of bare nanorod. This indicates that the oscillator strength, which is directly proportional to the radiative rate (*2*), is greatly enhanced for the QW exciton by a factor of more than 10, while the non-radiative rates remain almost same.

## S2. Numerical simulation of Fabry-Pérot cavity modes

The numerical simulations for Fabry-Pérot cavity modes of the nanorod cavities were performed using a commercial finite-difference-time-domain (FDTD) software (Lumerical FDTD solutions). A point dipole source with the wavelength range from 350 to 450 nm was used as an electromagnetic emitter. To reflect the polarization property of ZnO exciton, the dipole source with the polarization along the direction perpendicular to the long axis (*c*-axis) of nanorod was located inside the nanorod cavity (*3*). The mesh size was 2 nm for the calculations. For a comparison with the experimental data, the diameter and length of nanorod cavity lying on the SiO$_2$ layer were set to be 640 nm and 2.75 μm, respectively. All simulation boundaries were set as perfectly matched layers to prevent unphysical scattering at the edge of the simulation box. The resonance wavelengths for Fabry-Pérot cavity modes were investigated by an electric field intensity ($|E|^2$) spectrum from the nanorod end. The principal Fabry-Pérot resonances could be classified as two waveguide modes of HE$_{22}$ and HE$_{13}$ by investigating the cross-sectional electric field and Poynting vector distributions, as shown in Fig. S2. In order to check the field transmitted to the Si substrate, two-dimensional power monitor was placed at the interface of SiO$_2$/Si. The field intensity transmitted to the Si substrate was about 0.08% and 0.14% for HE$_{22}$ and HE$_{13}$ modes, respectively, indicating that the mode intensity absorbed to the Si substrate can be ignored. These results show that the Fabry-Pérot resonances in the nanorod cavity geometry are dominated by HE$_{22}$ and HE$_{13}$ waveguide modes.

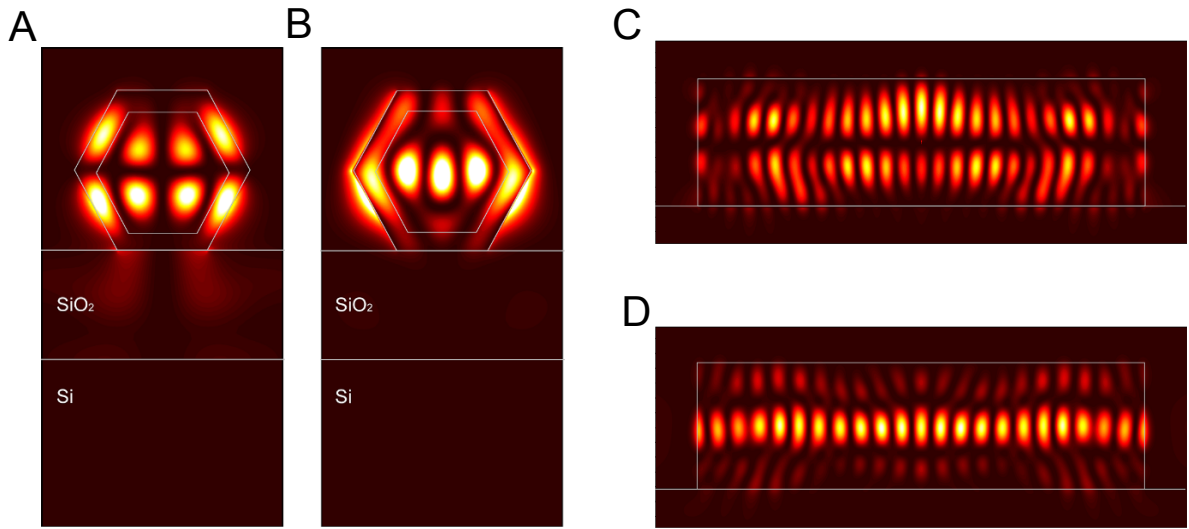

**Figure S2**. **Electric field distribution for HE$_{22}$ and HE$_{13}$ waveguide modes in the nanorod cavity geometry.** (**A**) Calculated $|E_y|^2$ distribution in the *x*–*y* plane for HE$_{22}$ mode at the wavelength of 387.7 nm. (**B**) Calculated $|E_x|^2$ distribution in the *x*–*y* plane for HE$_{13}$ mode at the wavelength of 392.2 nm. (**C**) Calculated $|E_y|^2$ distribution in the *y*–*z* plane for HE$_{22}$ mode at the wavelength of 387.7 nm. (**D**) Calculated $|E_x|^2$ distribution in the *y*–*z* plane for HE$_{13}$ mode at the wavelength of 392.2 nm.

## S3. Polariton dispersion curves with guided modes in QW nanorod cavity

The polariton dispersions of the nanorod cavity were calculated by a model for one-dimensional polaritonic dispersion (*4,5*). The polariton dispersion in the strong coupling regime can be expressed as (*6*)

$$n_{cav}^2 = \varepsilon(\omega, k) = \varepsilon_b + \frac{f}{\omega_0^2 - \omega^2 - i\omega\gamma} = \frac{c^2 k^2}{\omega^2}$$

where $\varepsilon_b$ is the background dielectric constant, $\omega_0$ are the exciton resonance frequency, $f$ is the oscillator strength, $\gamma$ is the damping constant and $n_{cav}$ is the refractive index of nanorod cavity. Also, the waveguide modes propagating along the cylindrical cavity geometry can be analytically calculated by the following eigenvalue equations (*4,7*).

HE$_{vm}$ and EH$_{vm}$ modes:

$$\left[\frac{J_v'(U)}{UJ_v(U)} + \frac{K_v'(W)}{WK_v(W)}\right]\left[\frac{J_v'(U)}{UJ_v(U)} + \frac{n_{air}^2}{n_{cav}^2}\frac{K_v'(W)}{WK_v(W)}\right] = \left(\frac{vk_z}{kn_{cav}}\right)^2 \left(\frac{V}{UW}\right)^4$$

TE$_{0m}$ modes:

$$\frac{J_v'(U)}{UJ_v(U)} + \frac{K_v'(W)}{WK_v(W)} = 0$$

TM$_{0m}$ modes:

$$\frac{J_v'(U)}{UJ_v(U)} + \frac{n_{air}^2 K_v'(W)}{n_{cav}^2 WK_v(W)} = 0$$

where the subscripts $v$ and $m$ denote the order and $m$-th root respectively, $J$ is the Bessel function of the first kind, $K$ is the modified Bessel function of the second kind. $U = r(k_0^2 n_{cav}^2 - k_z^2)^{1/2}$, $V = rk_0(n_{cav}^2 - n_{air}^2)^{1/2}$, and $W = r(k_z^2 - k_0^2 n_{air}^2(\lambda))^{1/2}$ are waveguide parameters, where $r$ is the cavity radius, $k_0$ is the free-space wave vector, $k_z$ is the propagation wave vector along the long axis ($z$) of nanorod and $\lambda$ is the free space wavelength. The polaritonic waveguide dispersions for HE, EH, TE, and TM modes can be calculated by substituting the refractive index $n_{cav}$ into above eigenvalue equations.

Figure S3 shows the polariton dispersion curves for waveguide modes in the nanorod cavity with the diameter of 640 nm. In this calculation, the exciton resonance energy of 3.32 eV was used to match with the experimental value measured at room temperature. The $\varepsilon_b$ (3.68) and $\hbar\gamma$ (0.7 meV) values were chosen from refs. (*8–10*). The experimentally measured resonance energies with equidistant momenta at the interval of $\pi/L_z$ were fitted to the calculated $E$–$k_z$ polariton dispersion by a parallel shift of the data points in the momentum axis. Note that the initial values of the equidistant momenta can be roughly estimated by simulating the field distribution of the Fabry-Pérot cavity mode since $k_z = n\pi/L_z$ where $n$ is the number of antinodes along the *z*-axis. The best fit for the measured polariton eigenmodes was obtained with the enhanced oscillator strength by a factor of 11 compared to the bulk ZnO (*11*) when the QW excitons are coupled to the $HE_{22}$ mode.

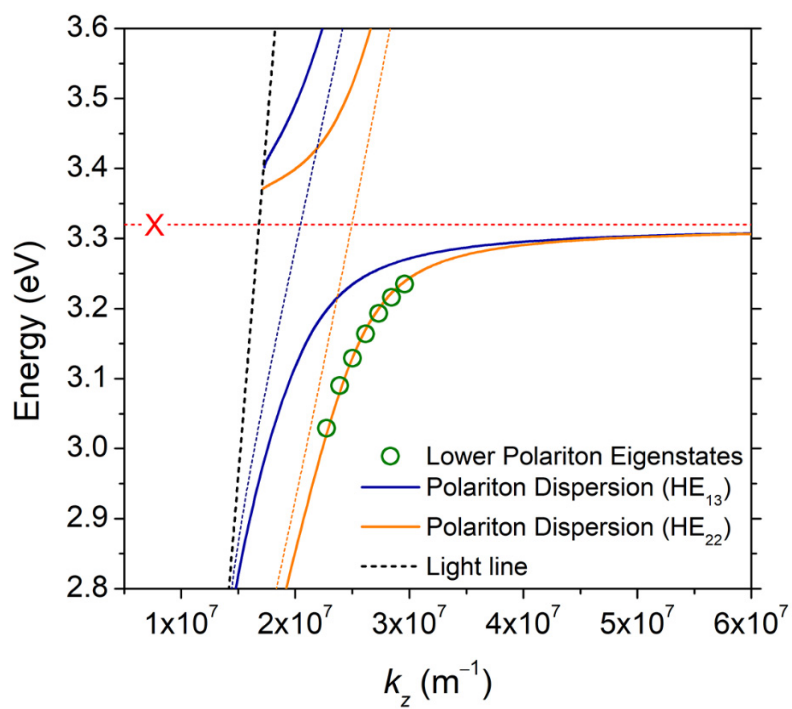

**Figure S3**. **Polariton dispersion curves.** Calculated polariton dispersion curves for $HE_{22}$ and $HE_{13}$ modes in the nanorod cavity with the diameter of 640 nm, showing that the best fit of the measured polariton eigenmodes was obtained by strong coupling of the QW exciton to the $HE_{22}$ mode. The black and red dotted-lines indicate the light line (air) and QW exciton energy at room temperature, respectively.

## S4. Mode volume, spatial overlap, and Rabi splitting energy in QW nanorod cavity

In the previous sections (Supplementary Materials S3), the Fabry-Pérot resonances including the exciton resonances in the nanorod cavity geometry were discussed to extract the polariton dispersion and coupling strength from the experimental data with the help of analytical and FDTD calculations. To further understand the strong coupling effect in the nanorod cavity geometry, we calculated the mode volume ($V_m$) of cavity and spatial overlap ($\Gamma_{overlap}$) between the field and QW excitons by using the following equations (*5,12*).

$$V_m = \frac{\int_V \varepsilon(r)|E(r)|^2 dr^3}{(\varepsilon(r)|E(r)|^2)_{max}} = \frac{L}{2} \times \frac{\int_A \varepsilon(r)|E(r)|^2 dr^2}{(\varepsilon(r)|E(r)|^2)_{max}}$$

$$\Gamma_{overalp} = \frac{\int_{active} \varepsilon(r)|E(r)|^2 dr^3}{\int_V \varepsilon(r)|E(r)|^2 dr^3}$$

where $V$ ($A$) is the simulation volume (area), $L$ is the length of nanorod cavity, $\varepsilon(r)$ is the dielectric constant, $|E(r)|^2$ is the intensity of electric field. The mode volume and spatial overlap for $HE_{22}$ mode were calculated by using the field distribution in the two-dimensional cross-section and averaged field intensity along the long axis (*z*) of the nanorod, as shown in the section S2. The mode volume for $HE_{22}$ mode was calculated to be about 0.104 μm$^{-3}$. The spatial overlap ($\Gamma_{overlap}$) between the field of $HE_{22}$ mode and QW excitons was 0.23 when considering the three-dimensional field distribution.

The large Rabi splitting energy (~370 meV) in the QW nanorod cavity was estimated from the experimental data and the theoretical model. We calculated the Rabi splitting energy, taking into account the enhanced oscillator strength of QW excitons, the mode volume, and the spatial overlap between the field and the QWs. The equation for Rabi splitting energy is given by (*5,12*):

$$\hbar\Omega = 2\sqrt{\frac{\hbar\omega_{LT} \cdot \hbar\omega_0 \cdot V_{QW}}{V_m} \cdot \Gamma_{overlap}}$$

where $\hbar\omega_{LT}$ is the longitudinal-transverse splitting of QW excitons, $\hbar\omega_0$ is the exciton resonance energy, $V_{QW}$ is the volume of 5 QWs, $V_m$ is the effective mode volume, and $\Gamma_{overlap}$ is the spatial overlap between the field and excitons. For bulks, the $\hbar\omega_{LT}$ and $f$ (oscillator strength) can be obtained by the relation of $\hbar\Omega = \sqrt{2(\hbar\omega_{LT})\cdot(\hbar\omega_0)} = \sqrt{(\hbar^2 e^2/\varepsilon_0\varepsilon_b m_0)\cdot(f/V)}$ (*12,13*), where $e$ is the electron charge, $m_0$ is the free electron mass, and $V$ is the volume of bulk semiconductor, indicating that the $\hbar\omega_{LT}$ is proportional to the oscillator strength. For A-exciton of bulk ZnO, the reported experimental values were that $\hbar\omega_{LT}$ = 2 meV and $f$ = 1.1×10$^5$ meV$^2$ (*11,14*). In contrast, the $\hbar\omega_{LT}$ value in the QW nanocavity was estimated to be ~22 meV from the polariton dispersion (section S3). Using $\hbar\omega_{LT}$ = 22 meV, $\hbar\omega_0$ = 3320 meV, $V_{QW}$ = 0.099 μm³, $V_m$ = 0.104 μm³, and $\Gamma_{overlap}$ = 0.23 and taking into account coupling with A and B excitons, the Rabi splitting energy was calculated to be 360 meV, which is in excellent agreement with the obtained value of 370 meV from the polariton dispersion (Supplementary Materials S3).

## S5. Angle-resolved emission spectra from QW nanorod and spatial coherence

A Fourier optics system was used to measure the angle-resolved emission spectra depending on the pump fluence (*15*). The Fourier plane of the microscope objective (40×, numerical aperture, NA = 0.6) was projected onto the entrance slit of a spectrometer (1200 grooves/mm grating) with a CCD detector (512 × 2048 pixels). As the long axis (*z*) of nanorod cavity was set to be parallel with the entrance slit of the spectrometer, we could obtain the phase correlation behavior of emission with different angles ($\theta$) in the *x–z* plane (*15*). The waveguided light emission is dominated at the two end facets of the nanorod, at which the light is emitted nearly spherically due to diffraction at the nanoscale apertures. The angle-resolved emission spectra show the phase-correlated interference patterns (Fig. S5A–C) because the coherent light emitted from the two end facets of the nanorod interferes with each other, resembling Young's double-slit experiments (Fig. S5D). According to Young's double slit experiments, the intensity at the Fourier plane can be described by the following equation (*15*).

$$I(\theta) = I_0 \cos^2\left(\pi \frac{L\sin(\theta)}{\lambda}\right)$$

where *I* is the intensity, $\lambda$ is the wavelength, and *L* is the cavity length. Extracting from Fig. S5C, we plotted the interference spectrum as a function of angle (sin($\theta$)) in Fig. S5E. Figure S5E shows that the interference equation well reproduces the measured interference pattern, indicating the spatial coherence of lasing emission from nanorod ends.

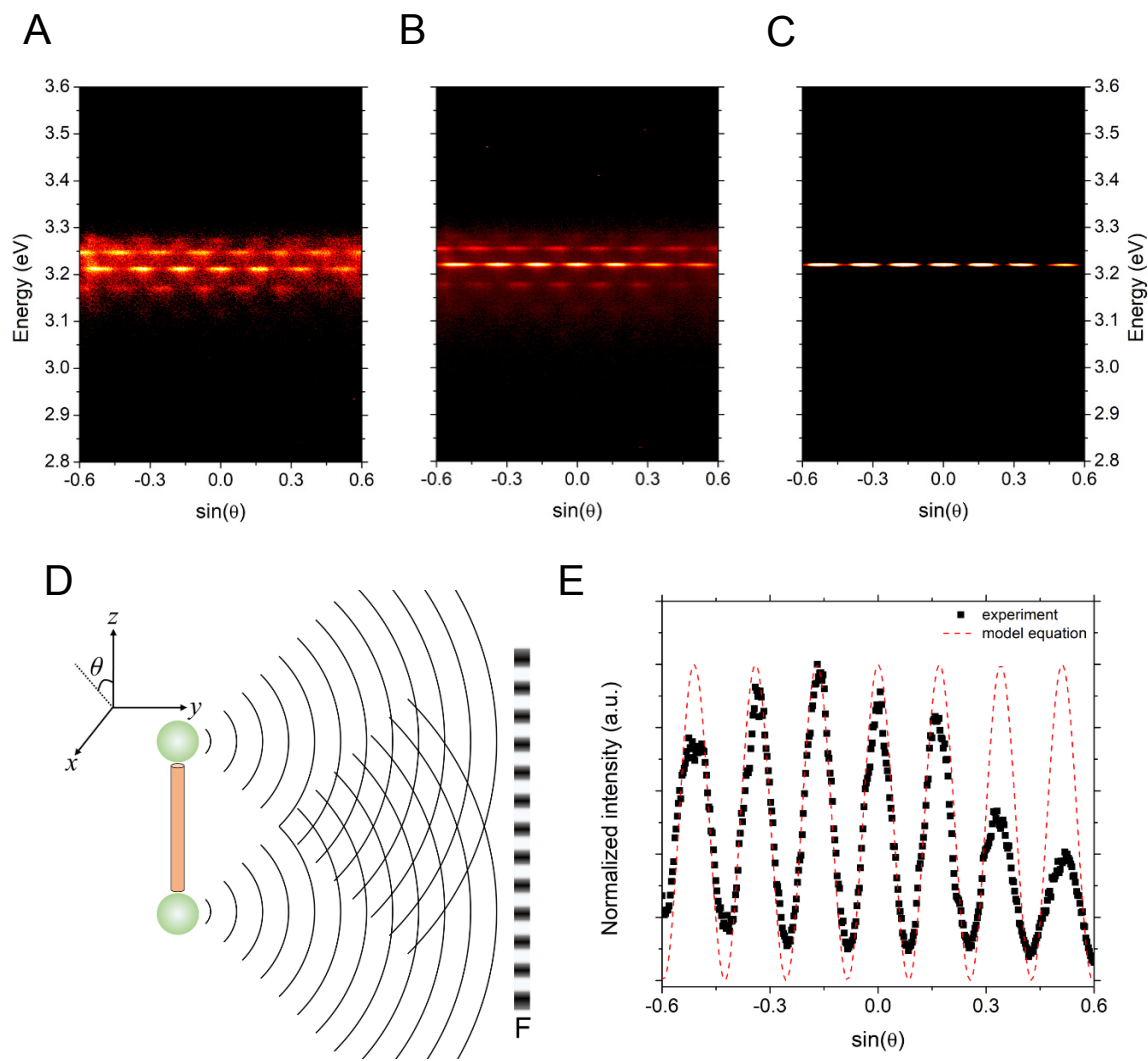

**Figure S5. Angle-resolved emission spectra from QW nanocavity.** (**A–C**) Angle-resolved emission spectra from QW nanorod at the pump fluence of $0.63P_{th}$ (A), $0.79P_{th}$ (B), and $P_{th}$ (C) ($P_{th}$: pump fluence at the lasing threshold), showing interference patterns. (**D**) Schematic illustration showing the far-field interference pattern due to the emission from the two end facets of the nanorod. (**E**) Interference spectrum of the lasing mode as a function of angle ($\sin(\theta)$).

## S6. Polariton dispersion above lasing threshold density

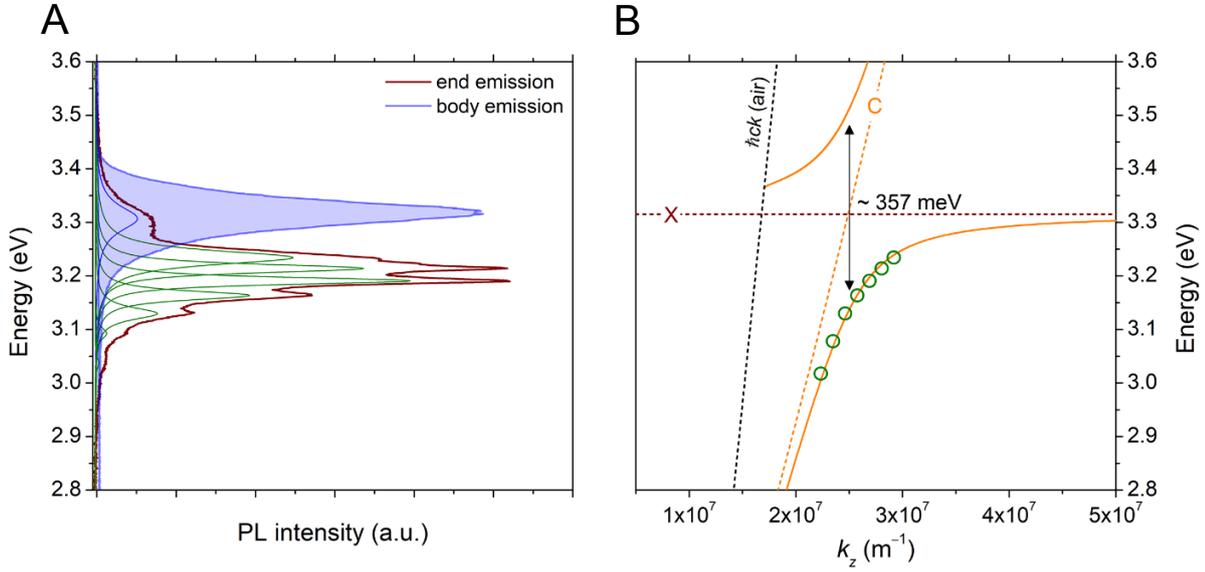

**Figure S6**. **Photoluminescence spectra and polariton dispersion at the exciton density above the lasing threshold.** (**A**) Photoluminescence spectra measured from the end and the body of QW nanorod under a excitation level of 2.2 kW/cm$^2$, corresponding to an electron-hole pair density of $3.5 \times 10^{17}$ cm$^{-3}$, which is exceeding the lasing threshold density. (**B**) $E$–$k_z$ polariton dispersion curve obtained from the end emission spectrum of Fig. S6A, showing the strong-coupling with Rabi splitting energy ($\hbar\Omega$) of 357 meV at the electron-hole pair density above the lasing threshold.

## S7. Statistical data for lasing thresholds

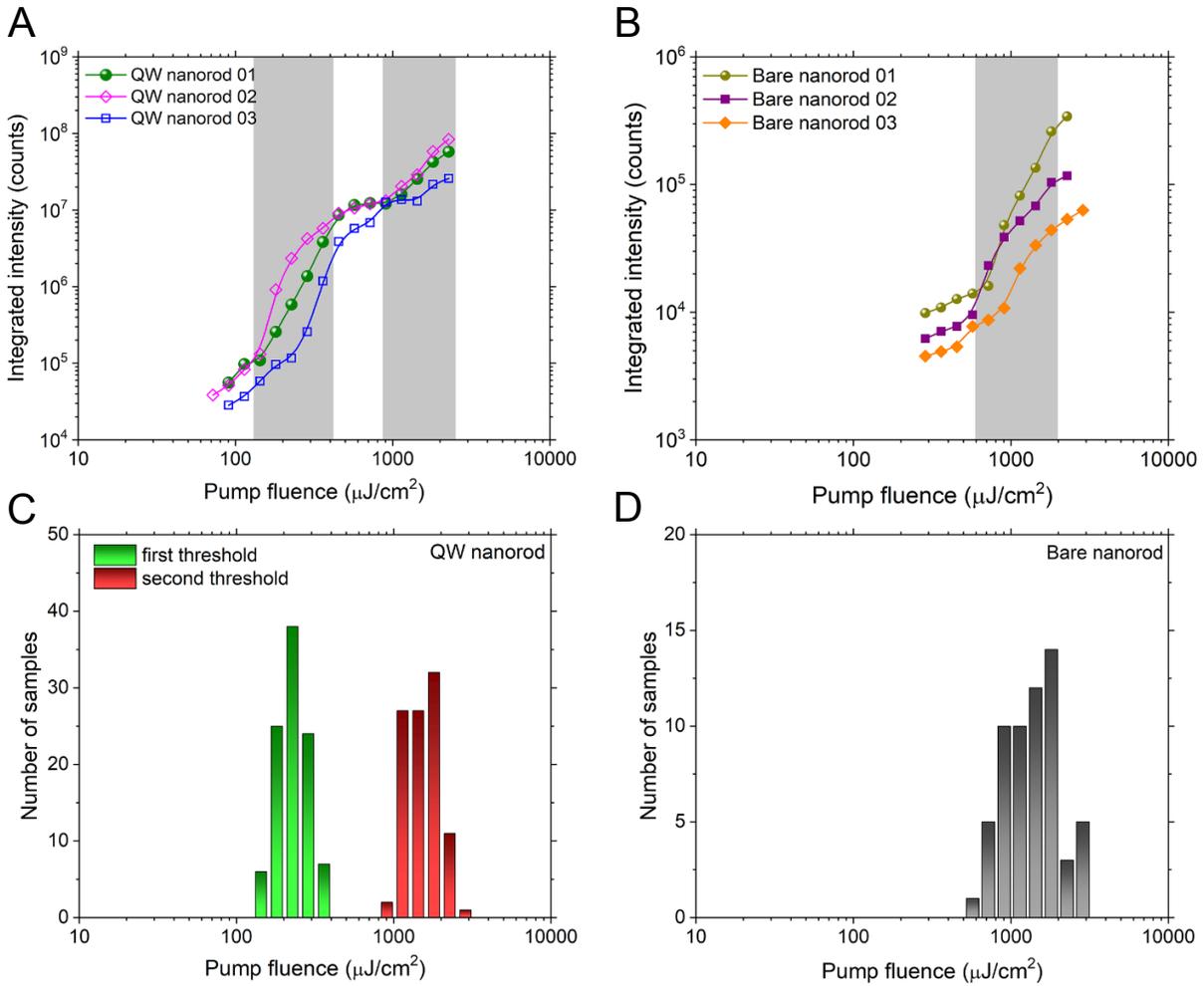

**Figure S7**. **Statistical data of the lasing threshold for QW and bare nanorods.** (**A, B**) Double-logarithmic plot of the integrated intensity as a function of the incident pump fluence for three different samples of QW (A) and bare (B) nanorods, showing the clear double threshold behavior in the QW nanorods. (**C, D**) Statistics for the lasing threshold from the QW (C) and the bare (D) nanorods, showing that the double threshold behavior for the QW nanorods was observed reproducibly. The statistics for the bare nanorods show that the lasing thresholds occur at the pump fluence exceeding the Mott density.

## S8. Room-temperature lasing spectra of bare ZnO nanorod cavity

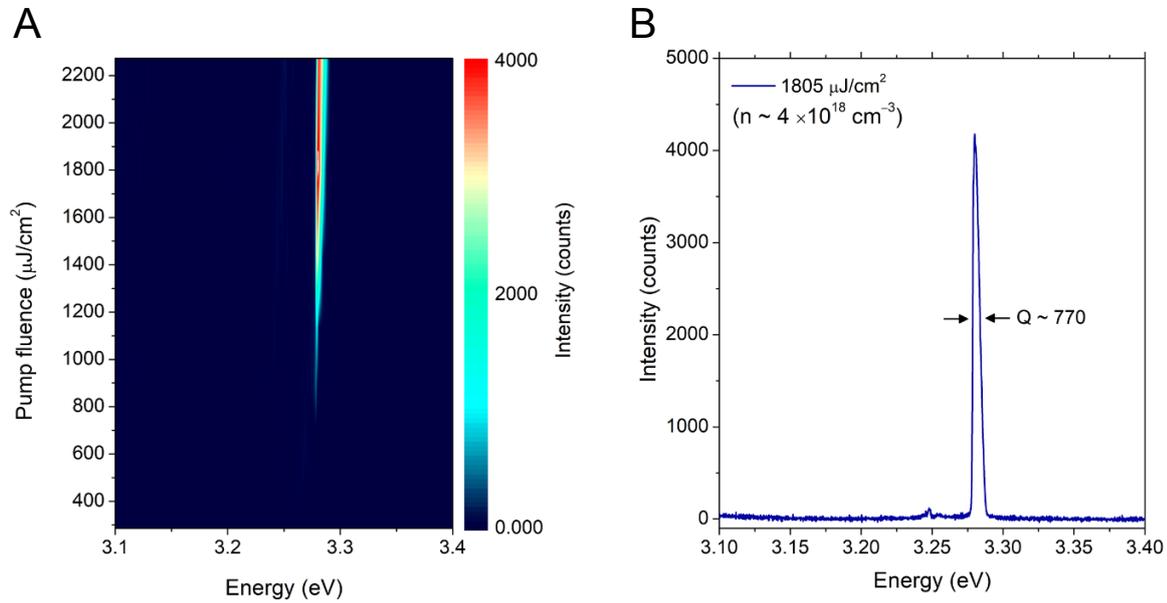

**Figure S8**. **Emission spectra of bare ZnO nanorod cavity depending on the pump flunece.** (**A**) Spectral map of bare ZnO nanorod (diameter of 600 nm, length of 2.53 μm) with increasing pump fluence at room temperature. The lasing threshold appears at the pump fluence of 905 μJ/cm$^2$. (**B**) Lasing emission spectrum at the pump fluence of 1805 μJ/cm$^2$, showing the lasing Q-factor of 770.

**S9. Estimation of electron-hole pair (EHP) density in bare and QW nanorods**

The lasing characteristics of QW nanorod at the room temperature were measured at the pumping wavelength of 355 nm (3.49 eV). As shown in the Fig. S9A, the $Zn_{0.9}Mg_{0.1}O$ with the optical band gap of ~3.55 eV, which is larger than the pumping energy, was used as the quantum barrier layer. At the given pump fluences, the EHP density was estimated by using the following equations (*16*).

$$I(t) = \frac{Fe^{-t^2/(2d^2)}}{\sqrt{2\pi}d}$$

$$\frac{dn(t)}{dt} = \frac{I(t)}{\hbar\omega D} - \frac{n(t)}{\tau_{decay}}$$

Here, *I(t)* is the time-dependent pump intensity for a Gaussian pulse, *F* is the fluence (in J/cm² per pulse), *d* is $1/\sqrt{8\ln 2}$ times the pulse width, $\hbar\omega$ is the photon energy, *D* is the diameter of the nanorod, and $\tau_{decay}$ is the carrier decay time. In case of bare nanorods, all photons entering to the nanorod are absorbed because the diameter of nanorod is much larger than the penetration depth of the pump pulse (~50 nm) (*16,17*). In the steady-state condition (*dn/dt* = 0), the EHP density per pulse was calculated by taking the carrier life time of 400 ps for the bulks (*16*). However, for the ZnO/ZnMgO heterostructure, the carriers could be excited in both the ZnO and ZnMgO layer and the excited carriers in the barrier layer would be relaxed into the ZnO well region in very fast time scale. The measured absorption coefficients of ZnO and $Zn_{0.9}Mg_{0.1}O$ films are shown in the Fig. S9B. The EHP density per pulse per QW is estimated from the number of photons per pulse multiplied by the net absorption of the QW nanorod system (including the ZnMgO layer) of 32.7% calculated by the transfer matrix method and divided by the number of QWs (5 pairs). Due to the thick barrier layer with the thickness of 10 nm, the effect of inter-well coupling was ignored. Note that the measured absorption coefficient was used to calculate the net light absorption in the total QW region.

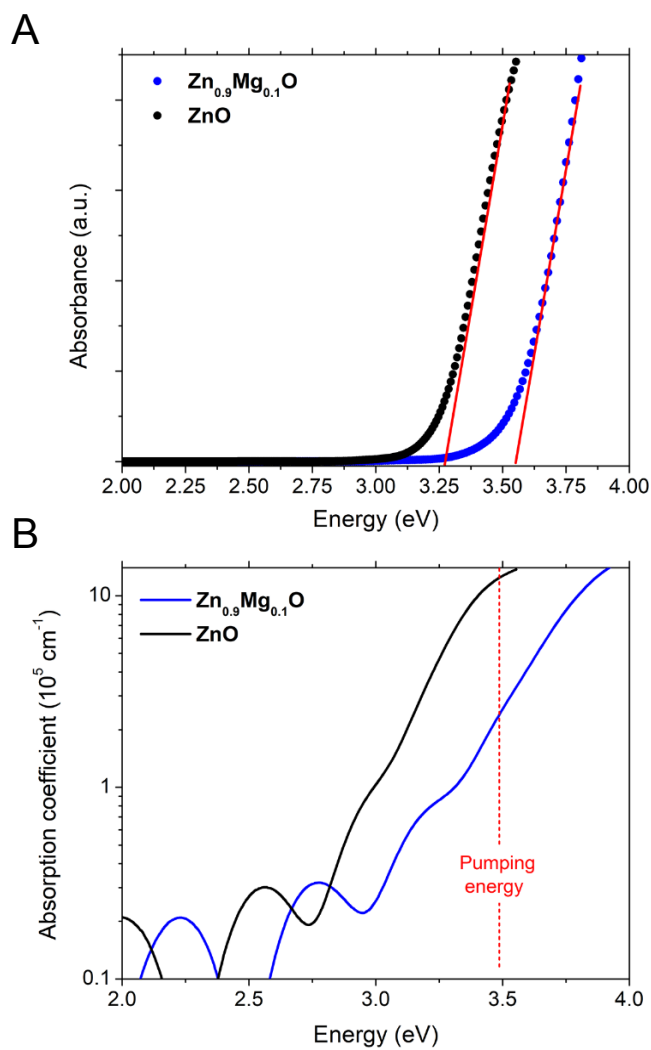

**Figure S9**. **Absorption spectra for ZnO and Zn$_{0.9}$Mg$_{0.1}$O layers.** (**A**) Absorbance spectra of ZnO and Zn$_{0.9}$Mg$_{0.1}$O, showing the room temperature optical bandgap of 3.26 and 3.55 eV for ZnO and Zn$_{0.9}$Mg$_{0.1}$O, respectively. (**B**) Absorption coefficients of the ZnO and Zn$_{0.9}$Mg$_{0.1}$O layers. The red dashed line indicates the pumping energy of 3.49 eV ($\lambda_{pump}$ = 355 nm).

# S10. Temperature-dependent lasing characteristics of bare ZnO nanorod

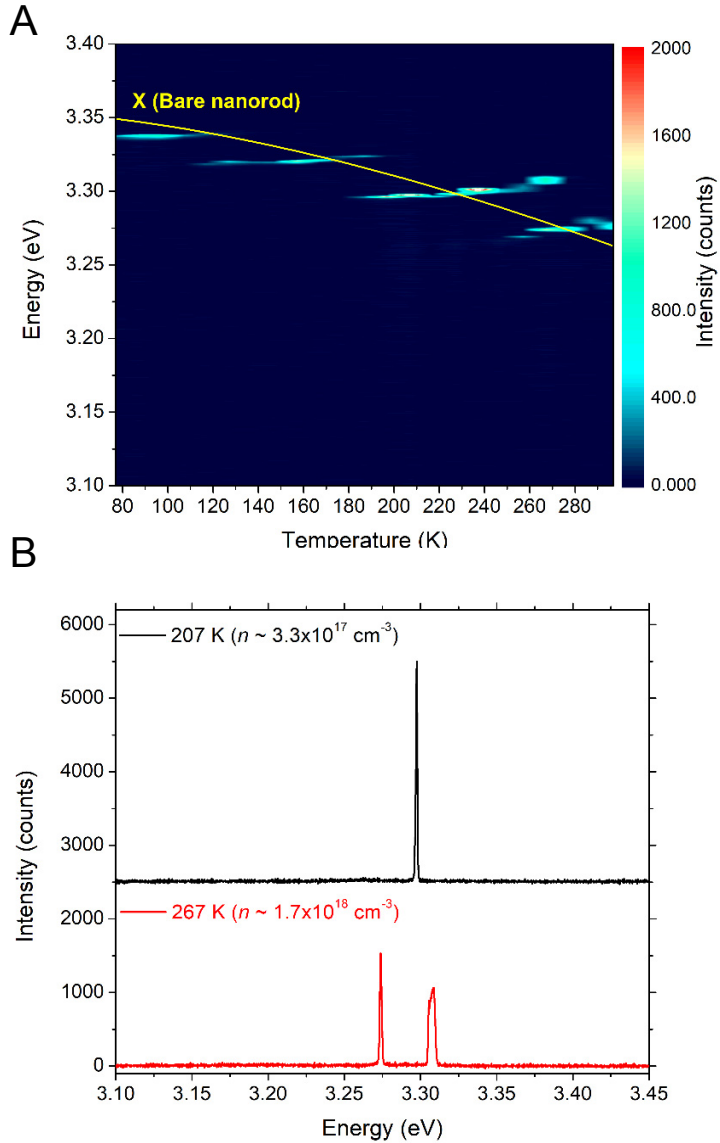

**Figure S10**. **Temperature-dependent lasing spectra of bare ZnO nanorod.** (**A**) Spectral map of temperature-dependent lasing emission of the bare nanorod measured at the pump fluence of $P_{th} \leq P \leq 1.26 P_{th}$. The lasing behaviors of bare nanorod in the temperature range from 197 to 297 K are significantly changed since the lasing threshold is rapidly increased as reaching to the room temperature. (**B**) Lasing emission spectra of the bare nanorod at the temperatures of 207 and 267 K. The lasing spectrum at 267 K shows the broad linewidth at near and above exciton energies, whereas the lasing at 207 K has a narrow linewidth along with the peak at sub-exciton energy. These temperature-dependent lasing behaviors in the bare nanorod indicate that the lasing mechanism is changed from the polariton to the photon lasing with increasing the temperature.